\documentclass[aps,prl,twocolumn,superscriptaddress]{revtex4}

\usepackage{graphicx}
\usepackage{amsmath}
\usepackage{amssymb}
\usepackage{exscale}
\usepackage[mathscr]{eucal}
\usepackage[sort&compress]{natbib}

\def\dbar{{\mathchar'26\mkern-12mu d}}

\begin{document}

\title{\large{A Complex Chemical Potential: Signature of Decay in a Bose-Einstein Condensate}}

\author{George E. Cragg}\affiliation{Department of Mechanical Engineering,}
\author{Arthur K. Kerman} \affiliation{Center for Theoretical
Physics, Laboratory for Nuclear Science and Department of Physics,
Massachusetts Institute of Technology, Cambridge, Massachusetts
02139}

\date{\today}

\begin{abstract}
We explore the zero-temperature statics of an atomic Bose-Einstein
condensate in which a Feshbach resonance creates a coupling to a
second condensate component of quasi-bound molecules.  Using a
variational procedure to find the equation of state, the
appearance of this binding is manifest in a collapsing ground
state, where only the molecular condensate is present up to some
critical density.  Further, an excited state is seen to reproduce
the usual low-density atomic condensate behavior in this system,
but the molecular component is found to produce an underlying
decay, quantified by the imaginary part of the chemical potential.
Most importantly, the unique decay rate dependencies on density
($\sim \rho ^{3/2}$) and on scattering length ($\sim a^{5/2}$) can
be measured in experimental tests of this theory.
\end{abstract}


\maketitle

Since the production of atomic Bose-Einstein condensates (BEC) in
the laboratory \cite{1,2}, many schemes have been proposed whereby
the experimentalist may control the interatomic interactions
governing the behavior of these gases \cite{3}.  One such proposal
involves a Feshbach resonance in which two atoms combine to form a
quasi-bound molecule \cite{4,5}.  This molecule is described as
the intermediate state or closed channel of the scattering
reaction as the constituent atoms in general have different spin
configurations in the bound state than in the scattering state.
Due to its dependence on the internal spin states, the energy
difference, or detuning between the scattered and bound states can
thus be tuned using the Zeeman effect in an external magnetic
field. As the binding energy of the molecular state is brought
close to the energy of the colliding atoms, the appearance of
these loosely bound molecules increases. Consequently, the
coupling between atoms and molecules acts to modify the effective
interatomic interactions. Near zero energy, these interactions are
described by the s-wave scattering length, which in turn can be
tuned by varying the external magnetic field. This degree of
control suggests that the negative scattering length of initially
unstable condensates may be tuned to positive values thereby
rendering the condensate stable.  We provide a many-body
variational description of such a scenario, specifically using the
case of $^{85}$Rb to compare our results with experiment
\cite{6,7}.  For a uniform system our findings reveal a collapsing
ground state and a decaying excited state, where the latter fits
the behavior seen from experiment. In the excited state, a complex
chemical potential is obtained in which the imaginary part
determines
the inverse decay time. \\
\indent Before embarking on the many-body analysis of the coupled
atom-molecule BEC, it is first necessary to review the relevant
two-body physics underlying the interatomic interactions.  It can
be seen that the two-body formalism provides a means by which the
many-body equations can be renormalized by replacing the
interaction strength by the s-wave scattering length as the
relevant parameter.  In two-atom scattering, there are a number of
different channels or outcomes of the scattering event,
corresponding to both the closed and open channels of the bound
and scattering states, respectively.  Because coupling will be
strongest to the highest lying molecular state below the continuum
of scattering states, we consider a two-channel model with
coupling between a single open and a single closed channel.  Since
the coupled channels analysis has been described in detail
elsewhere \cite{4}, we present only the main results of our
approach.  Using a separable form for the the two-body
pseudopotential we have,

\begin{equation} \label{1}
\langle \mathbf {k} \vert V \vert \mathbf{k'} \rangle = \lambda
v(\mathbf{k})v(\mathbf{k'}),
\end{equation}

\noindent where $\mathbf{k}$ is the momentum in the center of mass
and $\lambda$ is the interaction strength which is taken to be
negative, since we assume an attractive background throughout
\cite{8}. Using the separable potential with a molecular form
factor equal to $v(\mathbf{k})$, the coupled channels analysis
yields an integral equation for the two-body wavefunction, $\Psi
(\mathbf{k})$,

\begin{equation} \label{2}
\left(2\mathbf{k}^2 - E\right)\! \Psi (\mathbf{k}) + \left(
\lambda - \frac{\lambda ^2 \alpha ^2}{\epsilon - E} \right) \!
v(\mathbf{k})\! \int \dbar\,^3k' v(\mathbf{k'}) \Psi
(\mathbf{k'})=0,
\end{equation}

\noindent where $E$, $\lambda$ and $\epsilon$ are the center of
mass energy, interaction strength and detuning, respectively.
Here and throughout all energies are on the scale of $\hbar^2 /
2m$.  Equation (\ref{2}) is recognized as the usual single channel
two-body integral scattering equation \cite{9}, where the
molecular coupling has replaced the background interaction
strength, $\lambda$, with an energy dependent strength, $\lambda -
\lambda^2\alpha^2/(\epsilon - E)$. One striking feature of this
equation lies in the observation that for some nonzero coupling,
$g$, the interatomic interaction may be produced by a single
Feshbach resonance, for which the scattering equation becomes

\begin{equation} \label{3}
\left(2\mathbf{k}^2 - E\right)\! \Psi (\mathbf{k}) =
\frac{g^2}{\varepsilon - E}\, v(\mathbf{k})\! \int \dbar\,^3k'
v(\mathbf{k'}) \Psi (\mathbf{k'}).
\end{equation}

\noindent If the detuning, $\varepsilon$, is far enough away from
the two-body energy, $|\,\varepsilon|\! \gg \!|\,E|$, then the
effective interaction strength is independent of energy ($-g^2 /
(\varepsilon - E) \approx -g^2 / \varepsilon$) and can be chosen
to be equal to $\lambda$, the strength of the original
pseudopotential. Thus, the two-body background potential can be
modeled by a single Feshbach resonance with a detuning that is
very far from the system's energy.  It will be seen that this form
of the interatomic force results in a simplified expression for
the two-body interaction operator in the many-body Hamiltonian. \\
\indent Starting with the relationship between the scattering
length, $a$, and the $T$ matrix, $ 8\pi a = \langle \mathbf{k}| T
| \mathbf{k'} \rangle |_{\mathbf{k}=\mathbf{k'}=0}$, we can derive
the renormalization equation for the interaction strength
$\lambda$. The separable potential of Eq.(\ref{1}) indicates a
separable form of the $T$ matrix as well, $\langle \mathbf{k}| T |
\mathbf{k'} \rangle = v(\mathbf{k})tv(\mathbf{k'})$.  Thus, using
the Lippmann-Schwinger equation, $T = V + VGT$, where $G$ is the
free particle Green's function, we can solve for the $T$ matrix
and, consequently, the background scattering length, $a_{bg}$.
Subsequently, we take the zero range limit for the interatomic
potential in which we allow $v(\mathbf{k}) \longrightarrow 1$,
thus giving

\begin{equation} \label{4}
\frac{1}{8 \pi a_{bg}} = \frac{1}{\lambda} + \frac{1}{b},
\end{equation}

\noindent where the range, $b$, tends to zero as $\lambda
\longrightarrow 0^-$ in such a manner that $a_{bg}$ remains
negative and finite \cite{10}. As found from Eq.(\ref{2}), we
replace the background strength $\lambda$ by $\lambda - \lambda^2
\alpha^2 / (\epsilon - E)$ to include the effect of molecular
coupling. With $E=0$, we use this substitution to obtain the full
scattering length

\begin{equation} \label{5}
a \left( \epsilon \right) = \frac{\epsilon \, a_{bg}}{\epsilon + 8
\pi a_{bg} \alpha^2}.
\end{equation}

\noindent Alternatively, the full scattering length can be
expressed in terms of the external magnetic field, $B$, since the
detuning is given as $\epsilon = \gamma (B-B1)$, with $\gamma$
being the atomic species dependent proportionality and with $B1$
as the offset field. These two-body results will be used to
renormalize the many-body treatment. \\
\indent In the many-body analysis, we begin with a Hamiltonian
which includes the atomic kinetic energy, atom-atom collisions,
the molecular detuning energy and a coupling between atoms and
molecules.  However, according to the discussion behind
Eq.(\ref{3}), we can replace the usual two-body term,
$\hat{\psi}^4$, with an auxiliary field, $\hat{\chi}$, coupled to
the atomic field, $\hat{\psi}$, through a single Feshbach
resonance, thereby giving a simplified Hamiltonian which is
quadratic in the atomic field instead of quartic \cite{11,12}:

\begin{multline} \label{6}
   \hat{H} - \mu \hat{N} = \sum_{\alpha\beta}\hat{\psi}_{\alpha}^{\dag}
   \, ( p_{\beta}^2 - \mu ) \delta_{\alpha \beta}\, \hat{\psi}_{\beta} +
   (\epsilon - 2 \mu) \sum_\alpha \hat{\phi}_{\alpha}^{\dag} \hat{\phi}_{\alpha}\\
   +(\varepsilon - 2 \mu) \sum_\alpha \hat{\chi}_{\alpha}^{\dag}\hat{\chi}_{\alpha}
   + \frac {\lambda \alpha}{\sqrt{2}} \sum_{\alpha \beta \gamma}
   (\, \hat{\phi}_{\alpha}^{\dag} F_{\alpha \beta \gamma}
   \hat{\psi}_{\gamma} \hat{\psi}_{\beta} + H.c.) \\
   +\frac {g}{\sqrt{2}} \sum_{\alpha \beta \gamma}
   (\, \hat{\chi}_{\alpha}^{\dag} F_{\alpha \beta \gamma}
   \hat{\psi}_{\gamma} \hat{\psi}_{\beta} + H.c. ).
\end{multline}

In anticipation of the variational procedure to be carried out, a
term $-\mu \hat{N}$ has been added where the Lagrange multiplier
is identified as the chemical potential.  Furthermore, the atomic
field is coupled to both the molecular field, $\hat{\phi}$, and
the auxiliary field, $\hat{\chi}$, with coupling constants
($\lambda \alpha, g$) and detunings ($\epsilon, \varepsilon$)
defined by Eq.(\ref{2}) and Eq.(\ref{3}), respectively.  Since the
zero range limit is taken, we can, without loss of generality,
take the same molecular form factor, $F$, setting it equal to the
$v$ defined in the separable potential, Eq.(\ref{1}). As
discussed, the background atom-atom interaction is reproduced by
setting $g^2 / (\varepsilon - E) = -\lambda$ with
$|\,\varepsilon|\! \gg \!|\,E|$.  In the zero range limit
($\lambda \longrightarrow 0^-$), these conditions are satisfied if
we set $\varepsilon =
2/\lambda ^2$ and $g = \sqrt{-2/ \lambda}$.  \\
\indent For the zero-temperature analysis of the Hamiltonian
\cite{13}, we work in the Schr\"{o}dinger picture using a static,
``\,squeezed'' Gaussian trial wave functional given by

\begin{multline} \label{7}
\Psi[\psi, \phi, \chi]  =  N_\psi exp \left\{ -\sum_{\alpha \beta}
\delta \psi'_{\alpha} \frac{1}{4} G^{-1}_{\alpha \beta} \delta
\psi'_{\beta} \right\}  \\ N_\phi exp \left\{ -\sum_{\alpha}
\frac{1}{2} \delta \phi^{\prime \, 2}_{\alpha} \right\} N_\chi exp
\left\{ -\sum_{\alpha} \frac{1}{2} \delta \chi^{\prime \,
2}_{\alpha} \right\},
\end{multline}

\noindent where $N_\psi$, $N_\phi$ and $N_\chi$ are normalization
constants, and the width, $G$, is taken as real.  Denoting $\psi'$
as the atomic field with $\psi$ as its mean, the atomic field
fluctuations are given as $\delta \psi' = \psi' - \psi$, with the
analogs of the molecular fluctuations defined similarly. In the
momentum space representation of the uniform system, all fields
assume their mean values times a Dirac delta function, $\delta
(\mathbf{k})$, whereas the Gaussian width assumes the diagonal
form, $G(\mathbf{k})^{-1}$.  Taking the expectation value of Eq.
(\ref{6}) with respect to the trial functional of Eq. (\ref{7})
gives an expression for the pressure, $-P = \langle \hat{H} - \mu
\hat{N} \rangle / V$, in a volume $V$. Extremizing the pressure in
$G(\mathbf{k})$, and in the mean fields, $\psi$, $\phi$, and
$\chi$, gives a set of variational equations, which in momentum
space can be expressed as

\begin{equation} \label{8}
G(\mathbf{k}) = \frac{1}{2} \sqrt{\frac{k^2 - \mu + \beta}{k^2 -
\mu - \beta}},
\end{equation}

\begin{equation} \label{9}
(\beta - \mu) \psi = 0,
\end{equation}

\begin{equation} \label{10}
-\frac{\beta}{\lambda} - \frac{\beta \alpha^2}{\epsilon - 2 \mu} +
\int \dbar \, ^3 k D(\mathbf{k}) + \frac{1}{2} \psi^2 = 0,
\end{equation}

\noindent where we have defined $\beta = \alpha \phi + g \chi$.
Additionally, we denote the atomic density, $\langle
\hat{\psi}^\dag \hat{\psi} \rangle / V$, and the anomalous atomic
density, $\langle \hat{\psi} \, \hat{\psi} \rangle / V$, by $\int
\! R(\mathbf{k}) + \psi^2/2$ and $\int \! D(\mathbf{k}) +
\psi^2/2$, respectively, where the fluctuations $R(\mathbf{k})$
and $D(\mathbf{k})$ are related through the Gaussian width as

\begin{equation} \label{11}
R(\mathbf{k}) = \frac{1}{2} \left[ \frac{1}{4}\,G^{-1}(\mathbf{k})
+ G(\mathbf{k}) - 1 \right],
\end{equation}

\begin{equation} \label{12}
D(\mathbf{k}) = \frac{1}{2} \left[ \frac{1}{4}\,G^{-1}(\mathbf{k})
- G(\mathbf{k}) \right].
\end{equation}

Using Eqs. (\ref{8})-(\ref{12}), expressions for the energy
density, $u = \langle \hat{H} \rangle / V$, and the mass density,
$\rho$, are obtained,

\begin{multline} \label{13}
u = \int \dbar^{\,3} k \, k^2 R(\mathbf{k}) + \beta \frac{1}{2}
\psi^2 + \frac{\epsilon}{2} \frac{\alpha^2 \beta^2}{(\epsilon -
2\mu)^2}
\\ - \frac{\beta^2}{2 \lambda} - \frac{\alpha^2 \beta^2}{\epsilon
- 2\mu} + \beta \int \dbar^{\,3}k D(\mathbf{k}),
\end{multline}

\begin{equation} \label{14}
\rho = \int \dbar {\,^3} k R(\mathbf{k}) + \frac{1}{2} \, \psi^2 +
\frac{\alpha^2 \beta^2}{(\epsilon - 2\mu)^2}.
\end{equation}

\begin{figure}[h]
\includegraphics{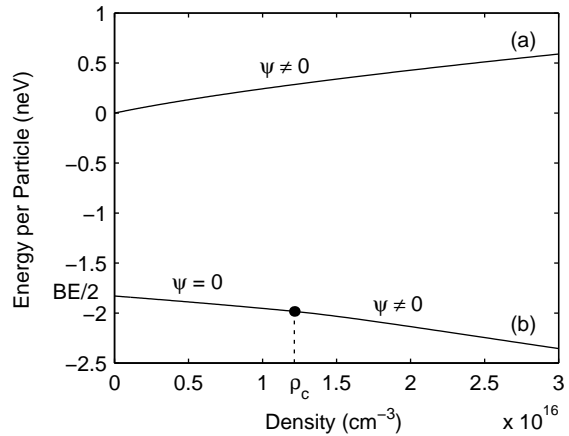}
\caption{The real part of the excited state (a) and the collapsing
ground state (b) are evaluated using the following numerical
values for $^{85}$Rb : applied magnetic field, $B=162.3$ G
\cite{6}, offset field, $B1=165.6$ G, coupling constant, $-8 \pi
a_{bg}^3 \alpha^2 = 330$, detuning proportionality, $\gamma = -30$
G$^{-1}$, and the respective background and full scattering
lengths, $a_{bg} = -450 \, a_0$ and $a(\epsilon) \approx 193 \,
a_0$, where $a_0$ is the Bohr radius \cite{14}. With these values,
we obtain a critical density, $\rho_c$, of $1.22$ x $10^{16}$
cm$^{-3}$ and a binding energy, BE, of $-3.7$ neV.  As the applied
magnetic field is decreased toward resonance, $\rho_c$ decreases,
thereby shortening the range of the $\psi = 0$ solution.  Tuning
the scattering length to negative values results in a collapsing
atomic-molecular condensate solution, with the usual low density
dependence, $e \sim 4 \pi a(\epsilon ) \rho$. \label{fig}}
\end{figure}

\noindent In the zero range $(b \longrightarrow 0, \lambda
\longrightarrow 0^-)$ limit, it can be seen from Eqs. (\ref{11})
and (\ref{12}) that the divergent terms in the integrals in $u$
have the behavior $\int \dbar^{\,3}k D(\mathbf{k}) \sim \beta /
2b$ and $\int \dbar^{\,3}k\, k^2 R(\mathbf{k}) \sim \beta^2 / 4b$.
Using Eq. (\ref{4}) relating $\lambda$ to the range, $b$, the
divergences cancel, resulting in an expression dependent only on
$a_{bg}$. This renormalization procedure is used in all solutions
given by Eqs. (\ref{8})-(\ref{14}).  \\
\indent We begin the analysis of these solutions by noticing that
for a nonzero $\beta$ in Eq. (\ref{8}), $\mu$ must be negative in
order to keep $G$ real.  Taking a negative $\mu$, we see from Eq.
(\ref{9}) that there are two solutions, with one given by a
vanishing atomic field ($\psi = 0$), and the other given by a
non-vanishing field and $\beta = \mu < 0$.  Considering the $\psi
= 0$ solution first, Eqs. (\ref{11}) and (\ref{4}) are substituted
into Eq. (\ref{10}), resulting in an expression that determines
$\mu$ in terms of $\beta$.  We find that this equation has a
solution for $\beta$ from zero only up to some critical value,
$\beta_c$.  From Eq. (\ref{14}), this range of beta corresponds to
a range of densities, starting from zero with $\beta$ and ending
at some critical value, $\rho_c$, corresponding to $\beta_c$.  As
shown in Fig. \ref{fig}, the plot of the energy per particle, $e =
u/\rho$, for this solution results in a downward curve starting at
half the molecular binding energy, BE/2, and stopping at $\rho_c$.
To find the remaining solution for $\rho > \rho_c$, we take $\beta
= \mu$ in Eq. (\ref{9}), and then use Eq. (\ref{10}) to solve for
$\psi^2/2$.  In this case, the energy per particle starts at
$\rho_c$, then continues downward for arbitrarily large $\rho$,
(Fig. \ref{fig}).  Thus, these two solutions comprise a single
curve of two pieces, recognized as the collapsing ground state of
the model Hamiltonian.  \\

\indent Thus far, we have not uncovered the solution corresponding
to the low density behavior, $e \sim 4\pi a(\epsilon) \rho$,
reported in the experiments for the positive scattering length
regime of $^{85}$Rb \cite{6}. Obtaining this solution requires a
closer examination of the density expression for $\beta = \mu$.
Using Eqs. (\ref{8}), (\ref{10})-(\ref{12}), the right hand side
of Eq. (\ref{14}) can be expressed in terms of the chemical
potential, $\mu$,

\begin{equation} \label{15}
\rho = \frac{\sqrt{2}}{3\pi^2}\left(-\mu\right)^{3/2} + \mu \left(
\frac{\alpha^2}{\epsilon - 2\mu} + \frac{1}{8\pi a_{bg}} \right) +
\frac{\alpha^2 \mu^2}{(\epsilon - 2\mu)^2}.
\end{equation}

Since Eq. (\ref{15}) has no solution for both $\mu, \rho > 0$, we
expand in $\rho$, which is constrained to be real and positive,

\begin{equation} \label{16}
\mu = 8 \pi a(\epsilon) \rho - i \frac{\sqrt{\pi}}{3} 256
a(\epsilon)^{5/2} \rho^{3/2} + \ldots,
\end{equation}

\noindent Thus, this expansion is seen to give the desired
low-density $4\pi a(\epsilon) \rho$ dependence of the energy per
particle, but becomes complex at higher order \cite{15}. \\
\indent At first glance, the complex-valued chemical potential
appears unphysical. However, the chemical potential can also play
the role of phase of the atomic field, $\psi$, as can most easily
be seen from the Heisenberg equation for $\psi$, $i\hbar \,
\partial \psi ^* /
\partial t = \langle [\hat{\psi}^\dag , \hat{H}] \rangle = - \beta
(t)^* \psi (t)$, in which the expectation value is carried out
using the trial wave-functional of Eq. (\ref{7}). Taking the time
dependence of the fields as $\psi (t) = \psi exp \, (-i \mu
t/\hbar)$ and $\beta (t) = \beta exp \,(-2i \mu t/\hbar)$, yields
the variational $\psi$ equation (\ref{9}).\\
\indent From this vantage point, the imaginary part of $\mu$ is
seen to be proportional to the inverse of the decay constant.  The
inspiration for such an interpretation arises in the context of
QED, where a complex action signifies decay of a constant, uniform
electric field \cite{16}.  Using the $^{85}$Rb parameters given in
Fig. (\ref{fig}) and taking $\sim 10^4$ atoms in a BEC cloud of
radius $25$ $\mu$m \cite{6}, a decay constant, $\tau$, of
approximately $14$ s is obtained to lowest order in $\rho$ ($ \tau
\sim 1/\rho^{3/2}$). The full solution of Equation (\ref{14})
leads to a seventh order equation in $\sqrt{-\mu}$, which, when
solved numerically, gives a decay constant of approximately $14.2$
s, thus confirming the validity of the expansion in Eq.
(\ref{16}). It must be pointed out that this is a novel decay
process inherent in the Hamiltonian describing only the two-body
interactions in the gas.  This result is in qualitative agreement
with the $10$ s decay reported in Ref. \cite{6}, where the
experiments were carried out in a regime where this decay process
dominates over the two and three-body inelastic processes
\cite{17}. As seen from Eq. (\ref{16}), this decay rate has novel
density and scattering length dependencies ($\sim
a(\epsilon)^{5/2} \rho^{3/2}$) which can be experimentally tested.
\\
\indent In summary, we have used a variational procedure in
extremizing a many-body Hamiltonian describing an atom-molecule
BEC coupled through a Feshbach resonance.  For positive scattering
lengths, we have found that the existence of the bound molecular
state results in a collapsing ground state and an excited state
with a complex-valued chemical potential.  By examining the
Heisenberg equation for the atomic field, the imaginary part of
the chemical potential is seen to physically correspond to the
inverse decay time.  Finally, this decay process was found to give
the dominant contribution in sufficiently dilute systems.
\begin{acknowledgments}
We thank Eddy Timmermans for many useful discussions.  This work
was supported by the MIT-Los Alamos Collaborative Research Grant
to Develop an Understanding of Bose-Einstein Condensates, contract
number 19442-001-99-35.
\end{acknowledgments}

\end{document}